\documentclass[a4paper,fleqn,usenatbib]{mnras} 
\usepackage{newtxtext,newtxmath}
\usepackage[T1]{fontenc}
\usepackage{ae,aecompl}
\usepackage{times,graphicx}	
\usepackage{amsmath}	
\usepackage{amssymb}	
\usepackage{color}

\usepackage{float}

\title[Simulations of gas sloshing in A1644]{Simulations of gas sloshing induced by a newly discovered gas poor substructure in galaxy cluster Abell 1644}

\author[Doubrawa et al.]{L. Doubrawa,$^{1,2}$\thanks{E-mail: lia.doubrawa@gmail.com}
R.~E.~G.~Machado,$^{2}$
T.~F.~Lagan\'a,$^{3}$
G.~B.~Lima Neto$^{1}$,
\newauthor R.~Monteiro-Oliveira,$^{1}$
and E.~S.~Cypriano$^{1}$
\\
$^{1}$Instituto de Astronomia, Geof\'isica e Ci\^encias Atmosf\'ericas, Universidade de S\~ao Paulo, Rua do Mat\~ao 1226, 05508-090 S\~ao Paulo, Brazil\\
$^{2}$Departamento Acad\^emico de F\'isica, Universidade Tecnol\'ogica Federal do Paran\'a, Rua Sete de Setembro 3165,80230-901 Curitiba, Brazil\\
$^{3}$N\'ucleo de Astrof\'\i sica, Universidade Cruzeiro do Sul / Universidade Cidade de S\~ao Paulo \\
	R. Galv\~ao Bueno 868, Liberdade, S\~ao Paulo, SP, 01506-000, Brazil\\
}

\date{Accepted 2020 April 16. Received 2020 April 10; in original form 2019 May 16}

\pubyear{2020}

\begin{document}
\label{firstpage}
\pagerange{\pageref{firstpage}--\pageref{lastpage}}
\maketitle

\begin{abstract}
Collision events lead to peculiar morphologies in the intracluster gas of galaxies clusters. That seems to be the case of Abell 1644, a nearby galaxy cluster, composed of three main structures: the southern cluster that exhibits a spiral-like morphology, A1644S; the northern cluster seen in X-ray observations, A1644N1; and the recently discovered substructure, A1644N2.  
By means of $N$-body hydrodynamical simulations, we attempt to reconstruct the dynamical history of this system. 
These simulations resulted in two specific scenarios:
(i) The collision between A1644S and A1644N2. Our best model has an inclination between the merger plane and the plane of the sky of $30^\circ$, and reaches the best morphology $1.6$ Gyr after the pericentric passage. At this instant A1644N2 is gas poor, becoming nearly undetectable in X-ray emission. This model shows a good agreement with observations;
(ii) The collision between A1644S and A1644N1. This approach did not give rise to results as satisfactory as the first scenario, due to great disturbances in density and mismatching temperature maps.
As a complementary study, we perform a three-cluster simulation using as base the best-fitting model to reproduce the current state of A1644 with the three main structures. This scenario presented a good agreement to the global morphology of the observations.
Thus, we find that the more likely scenario is a collision between A1644S and the newly discovered A1644N2, where A1644N1 may be present as long as it does not greatly interfere in the formation of the spiral feature. 
\end{abstract}


\section{Introduction}
Clusters of galaxies are dynamically young systems. In the hierarchical formation scenario, systems with smaller masses merge earlier, while more massive objects grow from accretion of minor structures \citep{White1991}.
These encounters between structures induce disturbances in the intracluster medium (ICM), in the form of shocks, cold fronts and other irregularities, that may become evident in the X-ray emission \citep{Markevitch2007}.

Cold fronts are understood as contact discontinuities in the ICM gas, which appear as discontinuities in temperature and density maps, but are continuous in pressure. They can be found in two varieties: in merging clusters such as A3667 \citep{Vikhlinin2001}, the Bullet Cluster \citep{Markevitch2002}, A2146 \citep{Russell2010, Russell2011}, among others \citep[see e.g.][]{Blanton2001, Blanton2011, Clarke2004, Ghizzardi2014, Ueda2019}; and also in apparently relaxed clusters in the form of a spiral that stems from the cluster cool core, which can reach out to large distances \citep{Rossetti2013, Walker2014, Walker2018, Douglass2018, Ichinohe2019}. Studies of nearby galaxy clusters performed by \cite{Lagana2010} showed that half of them have signs of spiral structures.

The gas sloshing mechanism, first proposed by \cite{Markevitch2001}, is understood as the origin of these spiral cold fronts, as a consequence of the gravitational impact during the pericentric passage in an off-axis collision. The cold gas of the cluster center starts to oscillate inside the main potential well, resulting in a spiral-like appearance of dense, cool, and low entropy gas that was removed from the cluster core. 

Details of the gas sloshing were studied through hydrodynamical simulations in \cite{Ascasibar2006} and \cite{ZuHone2010}, revealing that even a small gasless subcluster can induce the spiral feature after an off-axis passage, highlighting the gravitational basis of this phenomenon. Idealized simulations of clusters mergers are efficient to explain properties of the cold fronts, the influence of the initial parameters, and also to reproduce some individual objects \citep[e.g.][]{ZuHone2011, Roediger2011, Roediger2012, Johnson2012, Suzuki2013, Donnert2014, Machado2013, Machado2015, Sheardown2018, Sanders2020}.

Here, we continue this approach by performing numerical simulations in an attempt to recover the dynamical history of Abell $1644$, a nearby galaxy cluster at redshift $\Bar{z} = 0.0471 \pm 0.0002$ \citep{Rogerio2019}, that also exhibits the spiral pattern of the cold front.  
\cite{Tustin2001} presented the most complete sample of cluster members ($141$ galaxies), finding a lack of evidence for substructure in the galaxy distribution, which resulted in a mean redshift and velocity dispersion for the entire cluster. These conclusions were in disagreement with previous redshift studies \citep{Dressler1988} and observations from \textit{Einstein X-ray Observatory} \citep{Jones1984} due to the presence of a substructure. Subsequent \textit{XMM-Newton} observations of \cite{Reiprich2004} confirmed the bimodal distribution and revealed a tenuous ICM ($4$--$6$ keV) connecting the two clusters (a main cluster in the south, and a northern subcluster), whose X-ray peaks coincide with two of the brightest galaxies in the region. The deepest X-ray data available for A1644 are the \textit{Chandra} ACIS-I observation from \cite{Johnson2010}, who presented mass estimates for the two main structures, and a precise determination of the projected separation between clusters centers ($700$ kpc). The authors discuss the collision between A1644S and A1644N1 and associate the presence of only one cold front spiral to an early stage of gas sloshing. In a first approach, comparisons with simulations from \citep{Ascasibar2006} suggested that the system is seen 700 Myr after the pericentric passage.

Recent studies of A1644 include \cite{Xcop}, which presents radial profiles of thermodynamic properties of the intracluster medium and \cite{Lagana2019}, who presents 2D maps of the spatial distribution of temperature, pressure, entropy and metallicity.
New gravitational lensing analysis \citep{Rogerio2019} updated the mass estimates and revealed the presence of a third massive structure that could also be the responsible for the sloshing phenomenon. The line-of-sight (LOS) velocity for all the three clusters was also estimated, showing a low relative LOS velocity between them. This behavior may be expected when the substructures are near apocentre, so the motion with respect to the main cluster is small regardless of the inclination of the orbital plane. This result contradicts the first assumption of a recent collision.

\begin{figure}
\includegraphics[width=\columnwidth]{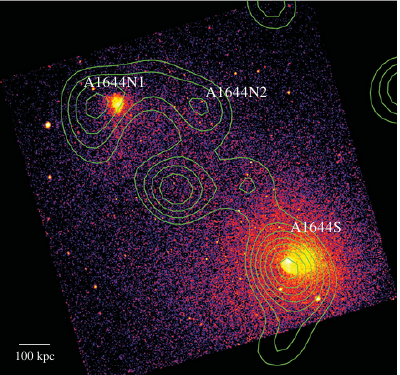}
\caption{\textit{Chandra} observation of A1644 overlaid with green contours showing the mass reconstruction from the gravitational lensing results of \protect\cite{Rogerio2019}. Here, we highlight the relevant structures of the galaxy cluster: the main southern cluster that exhibits the sloshing spiral, referred to as A1644S; the northern cluster, clearly seen in X-ray, also referred to as A1644N1; and the recently discovered structure, nearly undetectable in X-ray, referred to as A1644N2.}
\label{Fig:fig_01}
\end{figure}

Fig. \ref{Fig:fig_01} presents the three main structures of Abell 1644: A1644S, the main southern galaxy cluster that exhibits the spiral morphology of gas sloshing; A1644N1, the northern cluster, seen in the X-ray observations; and A1644N2, a nearly undetectable structure in X-ray, which was revealed by the gravitational lensing analysis. Green contours represent the mass distribution peaks of the significant structures recovered by the LensEnt2 code \citep{Marshall2002}, for more details see \cite{Rogerio2019}. 
This recently discovered structure has a virial radius of $r_{200}=0.86_{-0.66}^{+0.12}$ Mpc, virial mass of $M_{200}=0.76_{-0.75}^{+0.37} \times 10^{14}~{\rm M}_{\odot}$, and projected separation of $\sim531$ kpc to the A1644S mass centroid \citep{Rogerio2019}.

It is conceivable that the gas of the substructure is stripped in such circumstances of a galaxy cluster merger, making it difficult to identify. Following this hypothesis, we explore the possibility of an off-axis collision between A1644S and A1644N2 and present numerical simulations that explore this scenario. We also explore the role played by A1644N1.

This paper is organized as follows. Section \ref{Hydrodynamical} presents the simulation techniques and initial conditions. In Section \ref{Temporal} we discuss the global merger evolution of our preferred model and explore the parameter space of different possible collisions. In Section \ref{Comparison} we compare the simulation results to observations. In Section \ref{Collision6} we present an alternative model for the cluster dynamics, and a three-body simulation. Finally, in Section \ref{Summary} we summarize our results and present the conclusions. We assume a standard $\rm{\Lambda}$CDM cosmology with $\Omega_{\rm{\Lambda}}=0.73$, $\Omega_{M}=0.27$ and $H_0=70 h_{70}$ km s$^{-1}$ Mpc$^{-1}$. Errors quoted are in the $68\%$ confidence limit ($1\sigma$).

\section{Simulation setup}  \label{Hydrodynamical}
The goal of this work is to reproduce some of the morphological features in A1644 through $N$-body hydrodynamical simulations; in special, the spiral morphology of the southern cluster (also referred to as A1644S). Here, we perform idealized collisions between two galaxy clusters that give rise to the sloshing phenomenon. Even with simplifications, this approach allows the reconstruction of possible scenarios for the dynamical history of A1644. 

The gas of the ICM is described as an ideal adiabatic gas with $\gamma=5/3$. Cooling is neglected since its time-scale is longer than the time-scale of the collision. The individual gravitational contribution of galaxies is also neglected, since they represent only a small fraction of the cluster total mass \citep[e.g.][]{Lagana2008}. Star formation and active galactic nuclei (AGN) feedback are not considered. Magnetic field should not greatly affect the cluster global morphology, so it is also disregarded. As the spatial extent is relatively small, about $\sim3$ Mpc, the cosmological expansion is ignored. Simulations were carried out with the smoothed particle hydrodynamic (SPH) code {\sc gadget-2} \citep{Springel2005}, with smoothing length of $5$ kpc. The evolution of the system is followed for $5$ Gyr, and part of the output analysis made use of the yt-project tools \citep{Turk2011}.

\subsection{Density profiles}
The collision is represented by two spherically symmetric galaxy clusters initially in hydrostatic equilibrium, that were created following the procedure described in \cite{Machado2013} and \cite{Ruggiero2017}. Each cluster is composed of $2 \times10^6$ particles, equally divided into dark matter particles and gas particles.

The dark matter halo follows the \cite{Hernquist1990} density profile:
\begin{equation}
\rho_{\rm h}(r) = \frac{M_{\rm h}}{2\pi} \frac{r_{\rm h}}{r(r+r_{\rm h})^{3}} 
\end{equation} 
where $M_{\rm h}$ is the total dark matter mass, and $r_{\rm h}$ is a scale length. This profile resembles the NFW profile \citep{NFW1997}, except in the outer regions, with the advantage of having a finite total mass.

The adopted profile for the gas distribution is the \cite{Dehnen1993} density profile:
\begin{equation}
\rho_{\rm g}(r) = \frac{(3-\gamma)~M_{\rm g}}{4\pi} ~ \frac{r_{\rm g}}{r^{\gamma}(r+r_{\rm g})^{4-\gamma}} 
\end{equation}
where $M_{\rm g}$ is the total gas mass and $r_{\rm g}$ a scale length. When $\gamma=0$, the profile becomes similar to the $\beta$-model \citep{CF1976} commonly applied to represent the ICM of an undisturbed galaxy cluster, with a flat core (i.e. constant central density). When $\gamma=1$, the resulting profile is the Hernquist profile, which allows the presence of a dense cool core, with a steep density in the center, resulting in a suitable setup to the sloshing phenomenon.

\subsection{Initial conditions}
The gravitational lensing analysis presented in \cite{Rogerio2019} suggests that there are two potential candidates responsible for the sloshing phenomenon in A1644S: the northern cluster (A1644N1) and another massive cluster, nearly undetectable in X-ray observations, A1644N2. In this work we will mainly explore the possibility that the disturber is A1644N2. 

The southern cluster's total mass is always $M_{\rm a}=2.85\times10^{14}~{\rm M}_{\odot}$, its dark matter and gas scale lengths are $r_{\rm a,h}=250$ kpc and $r_{\rm a,g}=500$ kpc, with a gas fraction of $f_{\rm a,gas} = 0.15$ in all cases. Similarly, in the initial conditions, the total mass of A1644N2 is always in the ratio $1$:$2.4$ of $M_a$, with a dark matter scale length of $r_{\rm b,h}=200$ kpc and gas fraction $f_{\rm b,gas} = 0.10$ 
\footnote{An initial gas fraction of $0.1$ was chosen instead of a more usual $0.15$ due to the mass--gas fraction relation found observationally for lower mass clusters as presented in \cite{Lagana2013}.}
These values ensure that both the total mass and the temperature of the resulting objects are within the same order of magnitude as the observed clusters.

Initially the clusters are placed $3$ Mpc apart along the $x$-axis, that is, more than twice the combined virial radii of the clusters. This separation gives enough time for possible numerical transients to dissipate. A relative velocity $v_0$  along the $x$-axis is chosen, as well as an impact parameter $b_0$ along the $y$-axis.

In the search of the best model, we ran several simulations with different combinations of initial conditions. Table \ref{Tab:Tab1} displays the ones used in this study. The sample contains variations of four parameters: the initial relative velocity between clusters from $500$ to $900$ km/s; impact parameter from $600$ to $1000$ kpc; central density of A1644N2, with $r_{\rm b,g}$ varying from $150$ to $550$ kpc; and inclination of the orbital plane with respect to the plane of the sky from $0$ to $60 ^\circ$.

\begin{center}
\begin{table}
\caption{Initial condition parameters of the simulations of the collision between A1644S and A1644N2. The best model (B) is used as the central reference for the four sets of comparisons displayed in Fig.~\ref{Fig:fig_03}}
\centering
\begin{tabular}{cccccc}
\hline
Model & Label & $r_{\rm b,g}$ (kpc) & $v_0$ (km/s) & $b_0$ (kpc) & $i$ ($^\circ$)\\ 
\hline
A & b$600$  	   & $350$ & $700$ & $600$ & $30$ \\
B & b$800$   	   & $350$ & $700$ & $800$ & $30$ \\
C & b$1000$ 	   & $350$ & $700$ & $1000$& $30$ \\
D & v$300$  	   & $350$ & $500$ & $800$ & $30$ \\
B & v$600$  	   & $350$ & $700$ & $800$ & $30$ \\
E & v$900$         & $350$ & $900$ & $800$ & $30$ \\
F & ${\rm r_g}150$ & $150$ & $700$ & $800$ & $30$ \\
B & ${\rm r_g}350$ & $350$ & $700$ & $800$ & $30$ \\
G & ${\rm r_g}550$ & $550$ & $700$ & $800$ & $30$ \\
B$_{i0}$  & i$0$   & $350$ & $700$ & $800$ & $0$  \\
B         & i$30$  & $350$ & $700$ & $800$ & $30$ \\
B$_{i60}$ & i$60$  & $350$ & $700$ & $800$ & $60$ \\
\hline
\end{tabular}
\label{Tab:Tab1}
\end{table}
\end{center}

\section{Results}	\label{Temporal}
After several simulations, we found a best model that reproduces some of the desired morphological features of the system. This is Model B, which represents the collision between A1644S and A1644N2, with an impact parameter of $b=800$ kpc, and $v=700$ km/s. In this section we present the temporal evolution of the best model, and then discuss the effect of variations in the parameter space.

\begin{figure*}
\centering\includegraphics[width=\textwidth]{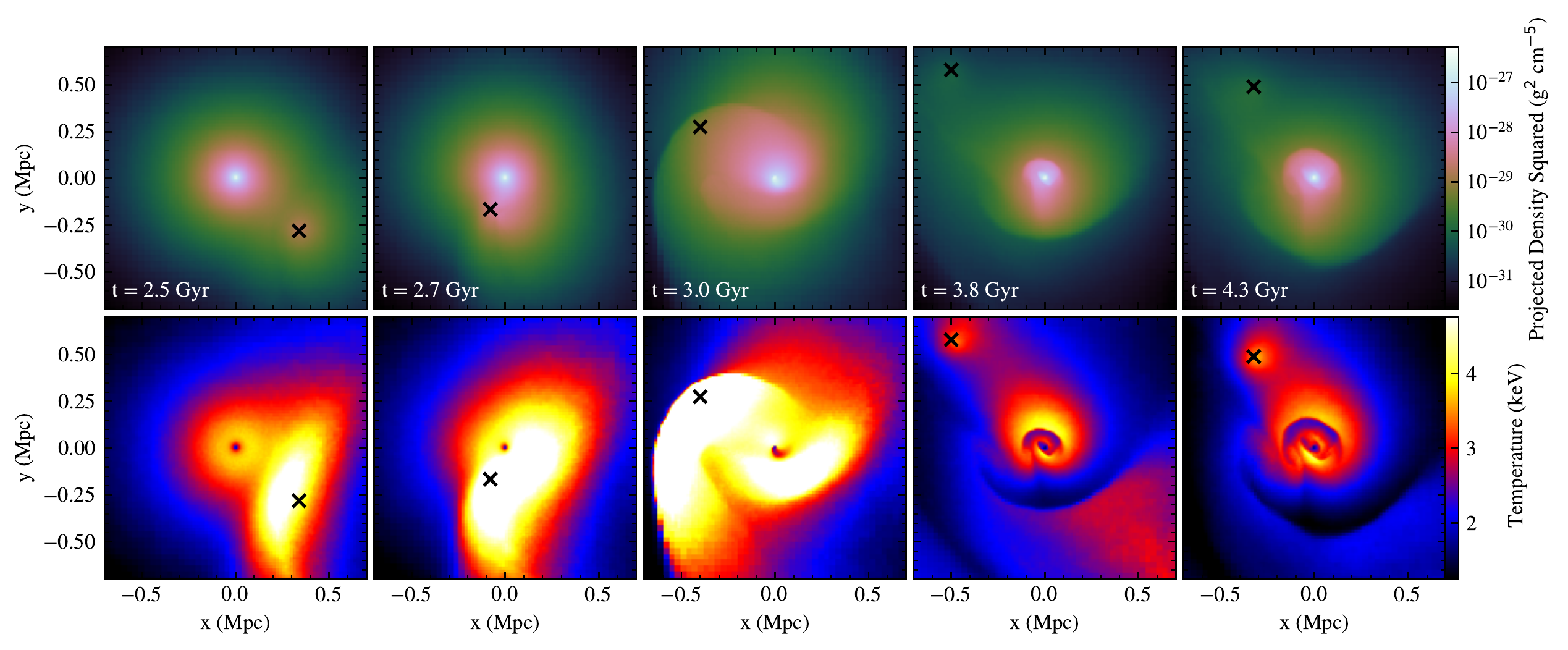}
\caption{Temporal evolution of the merger simulation between A1644S and A1644N2 (model B). These snapshots were chosen to highlight the beginning of the shock, the pericentric passage and the development of the sloshing phenomenon. Top panels: Projected density squared, with overlaid white contours representing the dark matter projected mass. Bottom panels: Emission-weighted temperature maps. The $\times$ symbols mark the position of the dark matter centroid of A1644N2.}
\label{Fig:fig_02}
\end{figure*}

\subsection{Temporal evolution} 
The temporal evolution of model B is displayed in Fig. \ref{Fig:fig_02} in a sequence of snapshots. The frames are $1100$ kpc wide, with non-uniform time intervals, in Gyr, that were chosen to highlight certain important moments of the simulation dynamic: the beginning of the shock, the pericentric passage and the development of the sloshing phenomenon. All the frames in Fig. \ref{Fig:fig_02} are projections with an inclination of $30 ^\circ$ in relation to the plane of the sky, discussed in details in Section \ref{inclination}. To allow a better visualization of the spiral of cold gas, we create density squared projections and emission-weighted temperature maps. The proxy employed for emission in galaxy cluster simulations is given by the free-free emission, $\propto n_e^2T^{1/2}$, as predominant at high energies.

In Fig. \ref{Fig:fig_02}, A1644N2 comes from the lower right corner of the frame towards A1644S, with an initial velocity of $700$ km/s, and at about $t=2.5$ Gyr, a shock wave starts to take form. At $t=2.7$ Gyr the pericentric passage takes place, with a distance between dark matter peaks of $180$ kpc. At this point, A1644N2 begins to lose its gas as it keeps following the trajectory until the apocentre at $t=3.8$ Gyr; meanwhile, A1644S starts to slosh due to the gravitational disturbance, developing the spiral feature. After reaching the maximum separation, A1644N2 returns to the southern cluster direction, and then, at $4.3$ Gyr, the desired gas morphology is reached, and also the desired projected separation of 550 kpc. At this moment, the presence of A1644N2 in the gas density map is hardly noticeable at all. In the temperature map, its temperature is roughly $4$\,keV.

\subsection{Exploration of the parameter space} \label{Exploration}
We explored different combinations of initial conditions, in order to constrain some of the collision parameters. The initial condition parameters are mainly: the masses of the clusters, their central gas concentration, the initial relative velocity and the impact parameter -- these may be chosen by construction. By the end of the simulation, the best model should satisfy simultaneously, to an approximate degree, the following criteria: the virial masses and radii for both structures, as measured in the \cite{Rogerio2019}; the general gas morphology, with a spiral of sloshed gas having an extent of approximately 200\,kpc; the projected separation between dark matter peaks of $\sim550$ kpc; and A1644N2 should have a low gas fraction, to be nearly undetectable in X-ray observations as presented by \cite{Johnson2010,Reiprich2004}.

Some of these constraints -- such as separation and virial masses -- are straightforwardly quantifiable. However, all the morphological comparisons -- between models and observations but also among models themselves -- are generally more qualitative. For example, the so-called `extent' of the spiral feature could be understood as the distance from the centre to the outermost edge of the cold front. The requirement that this size be roughly comparable to the size of the observed spiral was mostly used to rule out models where the spiral feature would be clearly unsuitable, either too large or too small. However, these judgements took into account also the shape of the spiral features and such decisions were based on qualitative visual inspection of the simulated maps. Therefore, in the following sections, when we discuss preferred models and compare agreements or shortcomings of different scenarios, it should be understood that the comparisons take into account both quantitative values but also qualitative morphological judgements.

Therefore, we chose the best morphological model by comparisons between the simulated temperature map and the observations \citep{Johnson2010,Lagana2019}, along with quantitative measurements of the profiles of thermodynamic properties \citep{Xcop}. A good qualitative agreement is achieved when the southern cluster develops a continuous round shape spiral after the collision event, and the disturber loses a good fraction of its gas in the pericentric passage, leaving a small trail of gas connecting the structures. Evident shock waves are not observed in the system surroundings. Temperatures of about $\sim$4.0\,keV are expected for the A1644S vicinity and $\sim$2\,kev for the core and the cold front. Detailed comparisons are presented in section \ref{Comparison}. 

The best model was obtained via trial and error tests with numerous attempts using different combinations of parameters. Here we present a small but systematic sample of models around this best-fitting model and discuss the main implications. Even with this search in space parameter, it is impossible to rule out the existence of another set of parameters that would also show acceptable results. Table \ref{Tab:Tab1} summarizes the sample models. 

\begin{figure*}
\includegraphics[width=2\columnwidth]{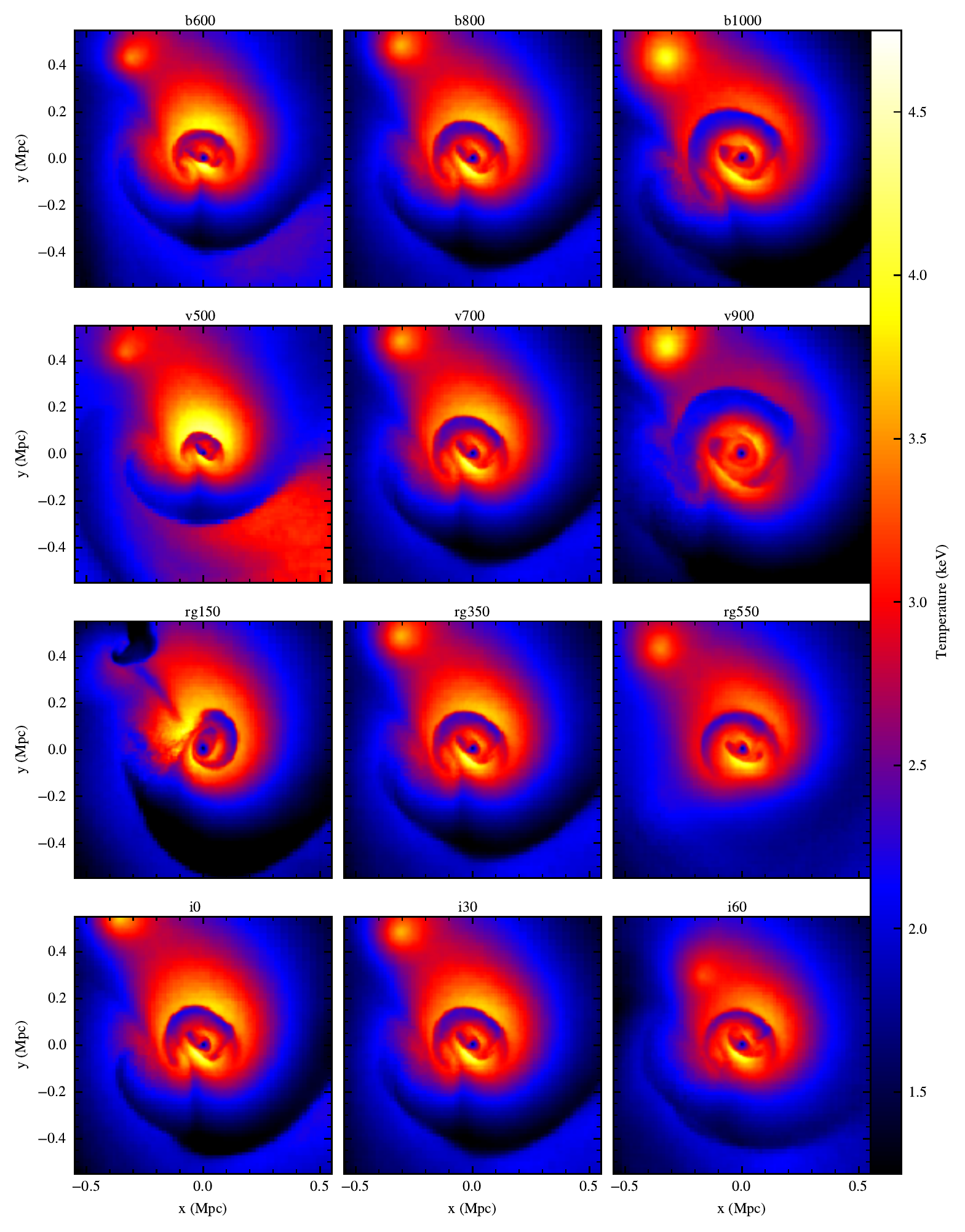}
\caption{Emission-weighted temperature maps for the merger simulation between A1644S and A1644N2 for different initial conditions. Each row displays three variations of one given parameter: (a) impact parameter; (b) initial relative velocity; (c) gas concentrations; (d) inclination. The central frame of each row is always the best model, to serve as a reference. The parameters corresponding to each label are given in Table~\ref{Tab:Tab1}.}
\label{Fig:fig_03}
\end{figure*}

The morphological comparison between emission-weighted temperature maps highlights the effects of varying each initial condition parameter separately. It also gives at least an approximate sense of the acceptable ranges of each parameter. Taking model B as fiducial, Fig. \ref{Fig:fig_03} shows variations in its surroundings. Each row displays variations of a given parameter: a) impact parameter; b) relative velocity; c) central density of A1644N2; d) inclination of the orbital plane with respect to the plane of the sky.
For each of these properties, $3$ variants are given, with one being the best model. With exception of row d) all the models are shown with same inclination, $i=30 ^\circ$, to allow a better comparison. All snapshots were chosen to represent the distances between peaks of 550 kpc without varying the inclination, this may lead to different instants of time for each combination of parameters. 

\subsubsection{Impact parameter}
Typical impact parameters in clusters mergers that present the sloshing phenomenon vary from dozens to hundreds of kpc \citep{ZuHone2011}. The models presented in row a) in Fig. \ref{Fig:fig_03} have impact parameters of $600$, $800$ and $1000$ kpc, respectively. The displacement, $b_0$, is along the $y$-axis in the initial condition, that is, when the $x$-axis separation is $3$ Mpc at $t=0$. 

Different displacements develop different spiral morphologies. In the case of small impact parameters, the disturbances in the gas morphology are larger, but the cold gas spiral does not develop properly. This occurs because the gravitational pull causes the core oscillation to become less circular, forming an excess of gas in a preferential direction. This is observed for $b=600$ kpc, whose excess is seen in the $x$ direction. On the other hand, with $b=1000$ kpc, the spiral is well formed, but A1644N2 does not lose enough gas to become nearly undetectable. At the instant of best separation, the spiral structure is somewhat larger than $200$ kpc.

The analysis allows to constrain a minimum separation $b_{\rm{min}}$, for the moment of closest approach between the clusters centers during the pericentric passage. The models presented in Fig. \ref{Fig:fig_03} a) have $b_{\rm{min}}=100$, $180$ and $260$ kpc. The presented discussion suggests a constrain of $b_{\rm{min}} > 180$ kpc, but if it is much larger, then A1644N2 would be able to retain too much of its gas.

\subsubsection{Initial relative velocity}
Row b) shows how different velocities can affect the spiral morphology, for $v=500$, $700$ and $900$ km/s. Given the cluster mass $M_{\rm a}$, the free-fall velocity of A1644N2 at $d_0=3$ Mpc would be approximately $640$ km/s, so we explore velocities in its surroundings. 

In $v=500$ km/s, the minimum separation during the pericentric passage is small ($\sim90$ kpc < $b_{\rm{min}}$), causing an increase in the ICM temperature (about $0.5$ keV higher when comparing to model B) and a large gas loss from A1644N2. This is necessary to become almost undetectable in X-rays, but the spiral of A1644S does not develop as it should. For $v=900$ km/s, at the instant of best separation, the cold gas spiral is substantially larger than what it should be, and the ICM temperature is quite smaller (about $0.5$ keV lower). At $v=700$ km/s the model shows a good overall morphology, a fair comparison to the ICM temperatures, as presented by \citep{Reiprich2004,Johnson2010} and the gas density of A1644N2 is low enough to become nearly undetectable in X-ray.  

\subsubsection{Central density}
In row c) of Fig. \ref{Fig:fig_03}, the effects of central density of gas are displayed. Models with $r_{\rm b,g}$ $=150$, $350$ and $550$ kpc are presented. These variations imply how bright the X-ray peak will be, i.e., it is prominent when the central density is high and unnoticeable otherwise. When A1644N2 is substantially denser, it remains more cohesive after the pericentric passage and does not lose enough gas to be nearly undetectable in X-ray imaging. One notices that the central gas concentration affects not only the final gas content of A1644N2 itself, but also the time scale of the orbit and the morphology of the spiral. As in model B the morphology is better represented, the preferred model is where $r_{\rm b,g}=350$ kpc.

\subsubsection{Inclination} \label{inclination}
Row d) in Fig. \ref{Fig:fig_03} shows the same instant in three different inclinations $i= 0$, $30$, and $60 ^\circ$. For higher inclinations, as $60^\circ$, the separation between the peaks of dark matter becomes smaller than $300$\,kpc. To have a good match between distances, earlier instants of time should be considered, resulting in gas morphologies that may be no longer acceptable. However, smaller inclinations are not strongly ruled out since the angular variation causes only a small variation in distance (from $550$ to $600$\,kpc). As an inclination of $30^\circ$ provides an adequate morphological agreement, this is adopted as the preferred value.

Although high inclinations would increase the LOS velocity of the system, excessively low inclinations would also provide LOS velocity near zero. In the \cite{Rogerio2019} analysis, measurements of the radial velocities among the substructures resulted in a low relative velocity. This is also what would be expected in clusters near their apocentre, where the overall velocity in relation to the principal cluster is low. However, for Model B, the system is seen approximately $0.5$\,Gyr after the maximum separation between the cores. To estimate the relative velocity in the LOS, we calculate a small displacement inside a time-step near the best instant of simulation in the $z$-axis. 
This rough approach gives a LOS velocity between A1644S/N2 of $\sim100$\,km/s, a result in agreement, within the error bars, to the estimated value of \cite{Rogerio2019}.

\section{Comparison with observations} \label{Comparison}

\subsection{Gravitational weak lensing results}
The initial conditions were created to satisfy the virial mass and virial radii, within $1\sigma$, computed by the gravitational weak lensing analysis presented in \cite{Rogerio2019}, namely $r_{200}=1.17_{-0.36}^{+0.16}$ and $0.86_{-0.66}^{+0.12}$ Mpc, $M_{200}=1.90_{-1.28}^{+0.89}$ and $0.76_{-0.75}^{+0.37} \times 10^{14}~{\rm M}_{\odot}$, respectively for A1644S and A1644N2. To ensure these values, we measured the spherically-averaged density profile, centering in the point of highest density, that is, the dark matter peak of the main cluster. We obtained a virial radius of $r_{200} = 1180$ kpc, and a resulting mass enclosed within $r_{200}$ of $M_{200} = 1.87 \times 10^{14}$ M$_{\odot}$. Similarly, when centered in A1644N2, the $r_{200}$ is $870$ kpc and have mass enclosed within $r_{200}$ of $M_{200} = 0.74 \times 10^{14}$ M$_{\odot}$. The virial masses are measured at the beginning of the simulation. By the time of the best instant of model B, that is $1.6$ Gyr after the central passage, some mass has been lost or redistributed. But as the gaseous part is the most affected due to the pericentric passage, the individual viral masses remain within the error bars.

\subsection{Residual X-ray surface brightness} \label{residuals}
From the outputs of the simulation, we can generate the mock X-ray surface brightness maps assuming thermal emission from the hot plasma, following the same procedure applied in \cite{Ruggiero2019}. 

\begin{figure}
\centering
\includegraphics[width=0.85\columnwidth]{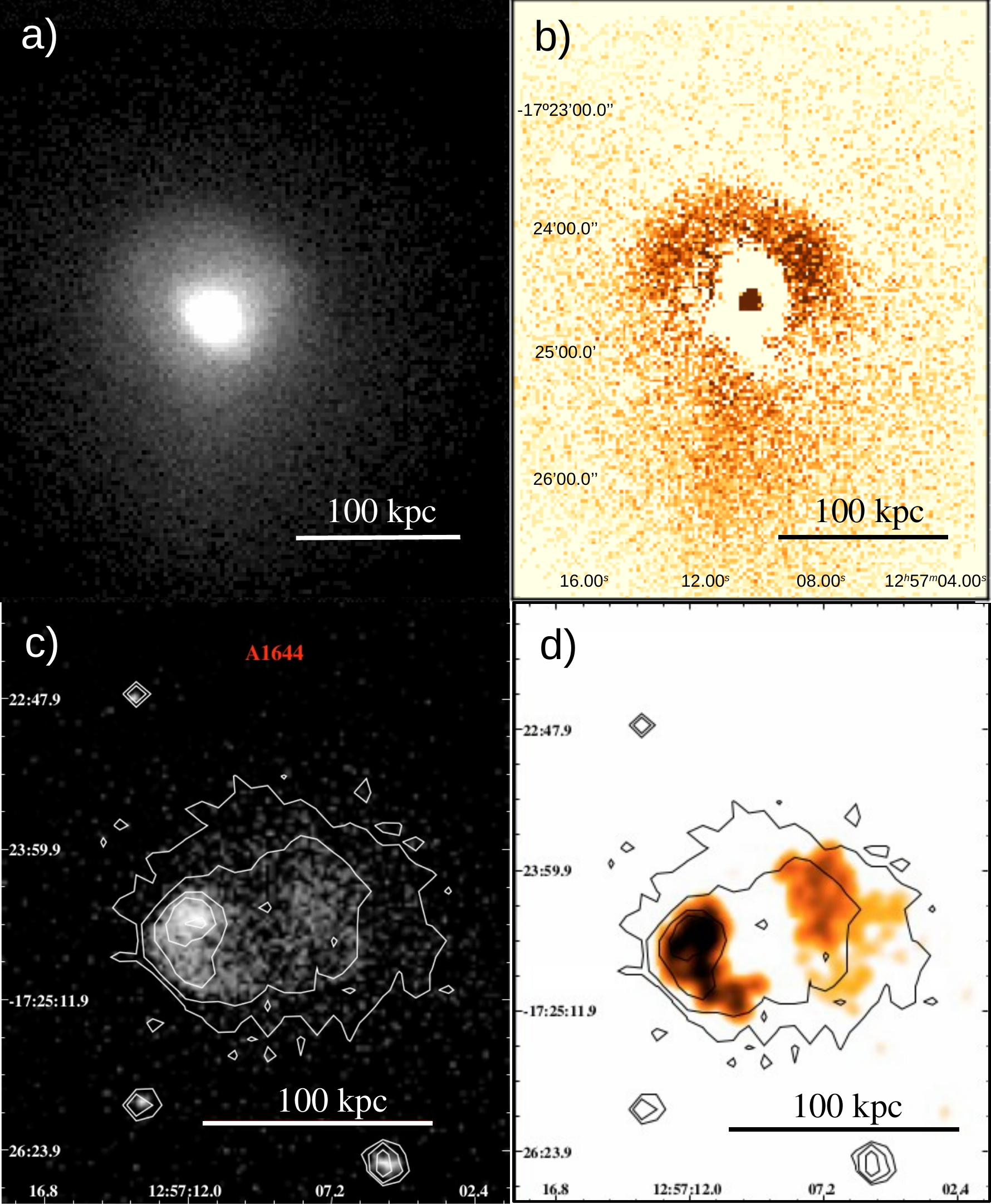}
\caption{Comparison between model B (collision between A1644S and A1644N2) at $t=4.3$ Gyr and \textit{XMM-Newton} observations \protect\citep[taken from][]{Lagana2010}: (a) Simulated X-ray surface brightness map; (b) Residuals from the subtraction of a fitted $\beta$-model; (c) Observed X-ray (d) Substructure map (residual+FOF algorithm\protect\footnote{The Friends-of-Friends (FOF) algorithm can be applied to highlight faint structures that are embedded in X-ray emission, by connecting neighboring pixels above a threshold. Fig. 3 in \protect\cite{Lagana2010} shows how this algorithm recovers the spiral pattern in Perseus cluster.} ) from the observational data.}
\label{Fig:fig_04}
\end{figure}

Using the python package pyXISM\footnote{http://hea-www.cfa.harvard.edu/$\sim$jzuhone/pyxsim/}, that follows the algorithm proposed by \cite{Biffi2012, Biffi2013,ZuHone2014} we simulate the X-ray observations from astrophysical sources. 
To this end, a constant metallicity of $0.3~Z_{\odot}$ is assumed and based on the simulated densities and temperatures of the gas, a photon sample is generated. These photons are then projected along the line of sight of the simulation.
Given the cluster coordinates, a foreground Galactic absorption model is applied \citep[with a neutral hydrogen column of $N_{\rm H} = 4.14 \times 10^{20}\,{\rm cm}^{-2}$;][]{Kalberla2005}. The photon list is then exported to be convolved with the \textit{Chandra} ACIS-I instrument response. The effective exposure time was $70$ ks and the energy range was $0.3-7.0$ keV for Fig. \ref{Fig:fig_04} and $0.5-2.5$ keV for Fig.\,\ref{Fig:fig_11}. The resulting images are shown in Fig. \ref{Fig:fig_04} a) for a zoom-in in A1644S and Fig. \ref{Fig:fig_11} a) for the three-body simulation (more details are presented in Sec.\ref{3bodies}). The energy bands were chosen to allow a fair comparison to \cite{Lagana2010} and \cite{Johnson2010}.

Using the {\sc ciao\footnote{asc.harvard.edu/ciao/}/sherpa\footnote{cxc.cfa.harvard.edu/sherpa/}} package, we have fitted a $2$D $\beta$-model onto the simulated spiral morphology, followed by the subtraction of this fit from its emission map. The result is a X-ray residual map. Fig. \ref{Fig:fig_04} displays the comparison of the simulated model with the observed X-ray excess. Even though a spiral morphology cannot be clearly discerned in the direct mock image, it can be seen as an excess in the residuals of the subtraction. Its general shape and orientation are roughly recovered, although the size of the spiral in the residuals seems overestimated.

\begin{figure}
\includegraphics[width=.95\columnwidth]{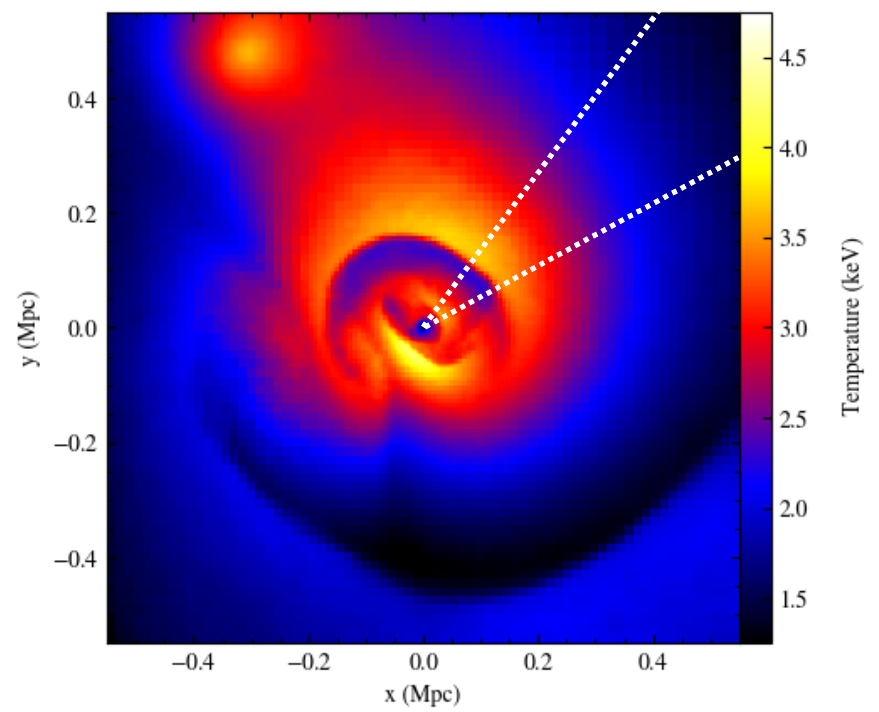}
\caption{Simulated emission-weighted temperature of model B (collision between A1644S and A1644N2) at $t = 4.3$ Gyr. The dashed lines represent, in projection, the cone within which the thermodynamic quantities of Fig.\protect\ref{Fig:fig_06} were measured.}
\label{Fig:fig_05}
\end{figure}

\begin{figure}
\includegraphics[width=0.95\columnwidth]{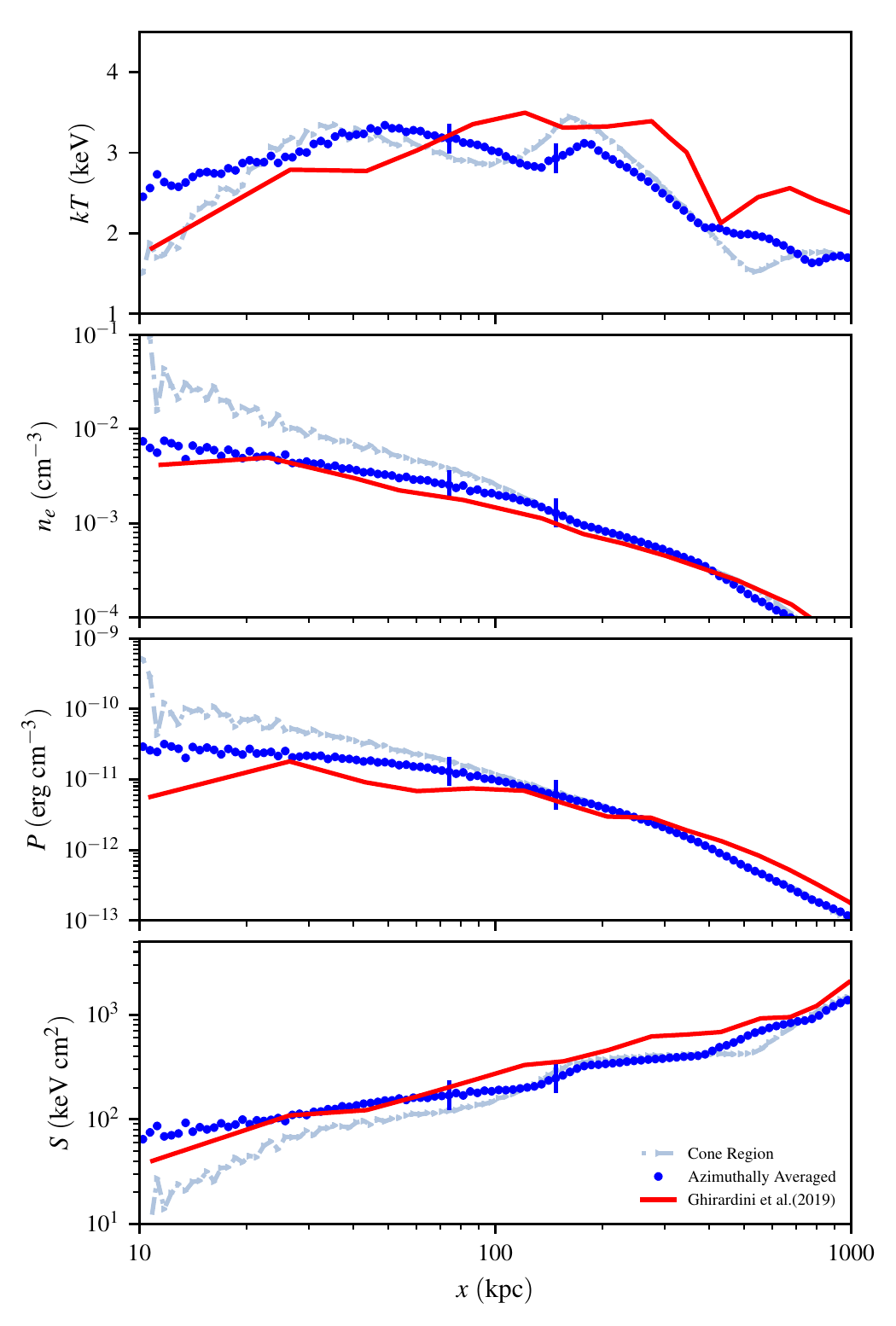}
\caption{Model B (collision between A1644S and A1644N2) at the best instant. The panels display, as light blue dash-dotted lines, the simulated temperature, electron number density, pressure and entropy profiles, measured within the cone region specified in Fig.~\protect\ref{Fig:fig_05}. Blue dots represents the same quantities within an azimuthally averaged radial profile. Red lines present observational results from \protect\cite{Xcop}. The vertical markers indicate the positions of the inner and outer faces of the cold front.}
\label{Fig:fig_06}
\end{figure}

\subsection{Radial profiles}
To better understand the spiral-like region, we compute some thermodynamic properties over an azimuthally radial profile alongside a conical region passing through the spiral-like structure displayed in Fig.~\ref{Fig:fig_05}. This is a commonly used technique when analyzing discontinuities in gas emissions that aims to highlight properties of the spiral that could be attenuated in a symmetrical profile \citep[e.g.][]{Kraft2004,ZuHone2010,Machado2015}. The resulting profiles of temperature, electron number density, pressure and entropy are summarized in Fig.~\ref{Fig:fig_06}. We adopt the usual proxy for the entropy, $S=kTn_e^{-2/3}$ \citep{Ponman2003}, in terms of electron number density $n_e$, temperature $T$ and Boltzmann constant $k$.

In Fig. \ref{Fig:fig_06} light blue dash-dotted lines represents the properties measured for the simulation output in the conical region, while blue and red lines are the azimuthally averaged radial profiles for the simulation output and the \cite{Xcop} observational results for A1644, respectively.
\cite{Xcop} computes their results rescaled by the $r_{500}$ and the thermodynamic quantity measured for each cluster. So, they were able to set global profiles for cool-core and non cool-core clusters. Here, we apply the $r_{500}$ and $M_{500}$ (utilized for the thermodynamic quantity calculation) derived from the gravitational lensing results \citep[ $r_{500}=0.76_{-0.23}^{+0.10}$ Mpc and $M_{500}=1.31_{-0.87}^{+0.60} \times 10^{14}~{\textrm M}_{\odot}$; derived from][]{Rogerio2019}. 
The comparison indicates a good quantitative agreement between the observational results and the best instant of model B.

This analysis shows that the spiral is composed by a cooler gas, about $\sim2$ keV, and also smaller entropy, $\sim20$ keV\,cm$^2$, than the gas in its immediate vicinity. In the outer face of the spiral a decrease in density is seen, which is not accompanied by a relevant discontinuity in pressure. This is an expected behavior for a cold front \citep{Markevitch2007}.

\subsection{A1644N2 gas mass}
It is possible to estimate an upper limit for the gas mass of A1644N2 from the X-ray counts in that direction. This result can then be directly compared with the gas mass measured from the simulations. 

In the \textit{Chandra} and \textit{XMM-Newton} X-ray observations, it is quite difficult to recognize A1644N2. Even though in both \cite{Reiprich2004, Johnson2010} analysis, the X-ray surface brightness contours hint at the presence of a substructure in its position. To have a gas mass estimate of A1644N2 we made use of the 52ks \textit{Chandra} X-ray observations and the {\sc Pimms}\footnote[6]{https://heasarc.gsfc.nasa.gov/docs/software/tools/pimms.html} simulation tool. In the X-ray observation, the count-rate inside $1$ arcmin in the A1644N2 position is $0.005^{+0.002}_{-0.003}$ count/sec. For the flux conversion, we assume a metallicity of $Z=0.2~{\rm Z}_\odot$, redshift of $z=0.047$ and $N_H$ absorption of $4.14 \times 10^{20}\,{\rm cm}^{-2}$. Given the observed flux, an upper bound to the gas mass of A1644N2 can be estimated, if a given density profile is assumed. From the simulation, we fitted a $\beta$-model to the density profile of model B with $r_c=9 \pm 5$ kpc and $\beta=0.6 \pm 0.1$. A spectral fit to the observations resulted in $\sim4$ keV, a temperature which is also consistent with the simulation result. The resulting estimated gas mass of A1644N2 was $M_{\rm gas}=4.6^{+12.8}_{-4.0} \times 10^{10}~{\rm M}_\odot$ inside $55$ kpc.

To compare this observational estimate with the simulations, we measured the spherically-averaged gas density profile, centering in the point of highest density of A1644N2. The resulting gas mass enclosed within $55$ kpc is $M_{\rm gas}=1.4 \times 10^{10}~{\rm M}_\odot$. This mass exceeds the observational estimate only by a factor of $\sim3$ of the central value and is well within the observational uncertainty. Since the simulation was not specially adjusted to meet this constraint a priori, this is a very good agreement. The simulation indicates that the gas mass fraction of A1644N2 should be roughly in the range 0.1--1 per cent, i.e.~far less than what would be expected for a non-perturbed $10^{14} {\rm M}_{\odot}$ cluster. Hence the term gas poor.

For a brief comparison between other models, we highlight Model F as the strongest case where the gas stripping was not enough to render A1644N2 nearly undetectable. Estimates of the gas mass enclosed within 55\,kpc resulted in $M_{\rm gas}= 8.5 \times 10^{10}~{\rm M}_\odot$. This gas mass is far from the central value of the observational result, by $4 \times 10^{10}~{\rm M}_\odot$, creating a gap that makes the subcluster detectable, even close to the estimated limit. The conclusion is that if the gas of the subcluster is more centrally concentrated, the stripping will not be as effective and it will be able to retain more of its original gas.

\vspace{-10pt}

\section {Simulations including A1644N1} \label{Collision6} 
In this section we explore another possibility for the formation of the spiral-like morphology: the collision between A1644S and A1644N1. This is presented in two distinct moments: the outgoing and incoming scenarios. After discussing the thermodynamic profiles and the collision implications, we present a three-body simulation that better reproduce the three main bodies of the galaxy cluster (A1644S, A1644N1 and A1644N2) and compare the resulting system to \textit{XMM-Newton} observations \citep{Lagana2019}.

We have previously presented our best model for the collision scenario of A1644, where the disturber was A1644N2. That scenario was achieved after several initial condition combinations, and later supported by the confirmation of the presence of a massive body through the gravitational lensing analysis \citep{Rogerio2019}. However, until recently, the most likely collision scenario would be with structures visible in the X-ray emission maps. In the case of A1644, previous studies revealed only one visible structure near to A1644S: A1644N1. 

For the reconstruction of the collision scenario between A1644S and A1644N1, another set of simulations was run. To reproduce some of the desired morphological features of the system, we describe A1644N1 as a spherically symmetric galaxy cluster with total mass of $M = 1.5\times10^{14}~{\rm M}_{\odot}$, dark matter and gas scale lengths of $r_{\rm a,h}=250$ kpc and $r_{\rm a,g}=450$ kpc, and gas fraction of $f_{\rm a,gas} = 0.15$. We maintained the same set of initial parameters from A1644S/N2 collision for the southern cluster.

Unlike N2, whose gas is described by a Dehnen profile, the initial conditions for N1 use a Hernquist profile due to the cool core observed by \cite{Johnson2010}. In the merger history, it is expected that after the pericentric passage, N1 will remain concentrated leaving only a small trail of gas. When the separation between dark matter peaks between A1644S/N1 is $700$ kpc, their X-ray emission peaks are comparable.

\begin{figure}
\includegraphics[width=\columnwidth]{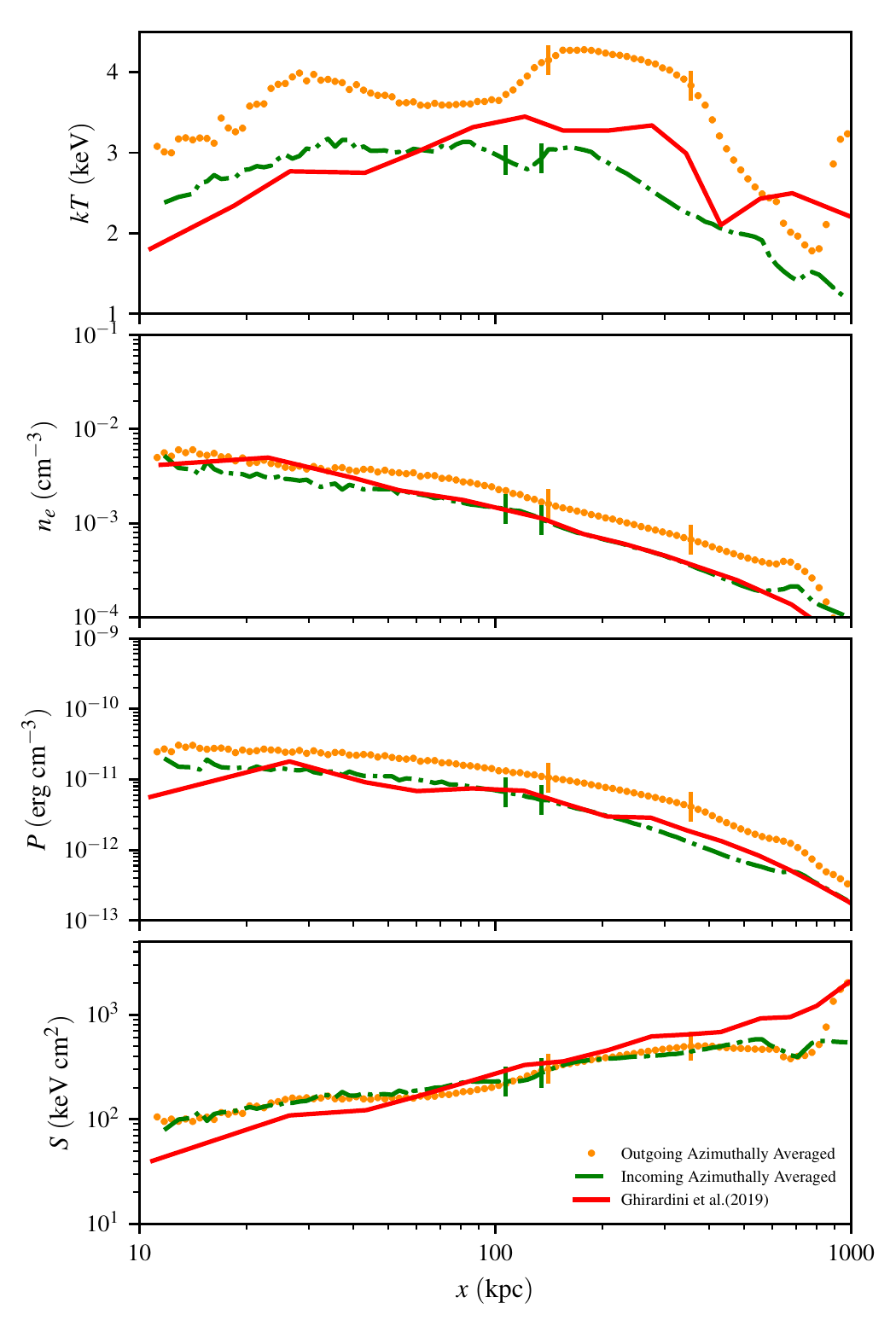}
\caption{Same as Fig.\protect\ref{Fig:fig_06} for the collision between A1644S and A1644N1 in the outgoing (orange dotted lines) and incoming scenarios (green dash-dotted lines). Red lines represent the \protect\cite{Xcop} results, applied for the $M_{500}$ and $r_{500}$ derived from \protect\cite{Rogerio2019}.}
\label{Fig:fig_07}
\end{figure}

As there is a preferential direction for the spiral morphology, the set up of this simulation is similar to the collision with A1644N2. The simulations starts at $t=0$, and is followed for $5$ Gyr. A1644N1 comes from the right corner, with an initial separation along the $x$-axis of $3$ Mpc, initial impact parameter, $b=700$ Gyr, and relative velocity, $v=1000$ km/s. 
After the pericentric passage, the system passes twice by best separation of $700$ kpc: once, in an outgoing scenario, right after the closest approach; and again, in the incoming scenario, after reaching the apocentre when the subcluster starts to fall back in the direction of the main cluster. 

We discuss, for these two regimes, the implications of a recent/older collision in temperature and density distributions, and also the thermodynamic properties in a radial profile.

\subsection{Outgoing scenario}  \label{outgoing}
The first best instant of the spiral morphology is at $t=3.3$ Gyr, that is $0.5$ Gyr after the pericentric passage, is presented in Fig. \ref{Fig:fig_11} in the ``Outgoing Scenario'' column, as mock X-ray emission with black contours representing the projected dark matter mass in the top row, temperature map in the middle and entropy in the bottom. 
In this scenario, large and excessively hot ($\sim6$ keV) shock waves are present. The spiral in A1644S is quite evident in the density map. Notably, there is a considerable symmetry between the two systems in this case: the perturber itself (A1644N1) develops a cool spiral of its own, comparable in size and importance to the southern spiral. This is not visible in the X-ray data. The twin spiral morphology that arises is one of the reasons to rule out this model.
 
The same thermodynamic properties presented in Fig.~\ref{Fig:fig_06} are measured for this model over an azimuthally averaged radial profile. The resulting profile is displayed in Fig.~\ref{Fig:fig_07} as dotted orange lines. A rough agreement is seen in all quantities, despite the large shock waves. Their presence is highlighted in the temperature profile, where temperatures close to $4.5$ keV are measured. Density, entropy and pressure shows some agreement at the spiral surroundings. 
The pressure profile shows a continuity, as expected for cold fronts. The simulation was tailored to have a good agreement (within $1\sigma$) with the $M_{200}$ given by the gravitational lensing analyses, with the separation between peaks of 700 kpc obtained to an inclination in relation to the sky of $14^\circ$ (value also consistent with the expected low LOS). 

\vspace{-10pt}

\subsection{Incoming scenario}  \label{incoming}
The second best instant in morphology occurs at $t=4.4$ Gyr, $1.6$ Gyr after the pericentric passage, when A1644N1 is returning after reaching the apocentre (Fig. \ref{Fig:fig_11}), in the ``Incoming Scenario'' column. As in the earlier scenario, the separation of 700 kpc is obtained with an inclination in relation to the sky of $14 ^\circ$, and the virial masses are tailored to have a good agreement with \cite{Rogerio2019} (within $1\sigma$). 
At this time, the shock waves associated with the pericentric passage of A1644N1 have advanced further into the cluster outskirts, making the ICM colder than in the outgoing scenario. Fig. \ref{Fig:fig_07} displays the thermodynamic properties for this instant as dash-dotted green lines. As in the outgoing scenario, the measurements shown an approximate agreement of the properties at the spiral structure surroundings.
The ICM temperature between the A1644S and A1644N1 is about $\sim2$ keV colder than in the observations \citep{Lagana2019}. Moreover, the extension of the spiral morphology is overestimated, that is, it has a length of $\sim300$ kpc.

\begin{figure}
\includegraphics[width=\columnwidth]{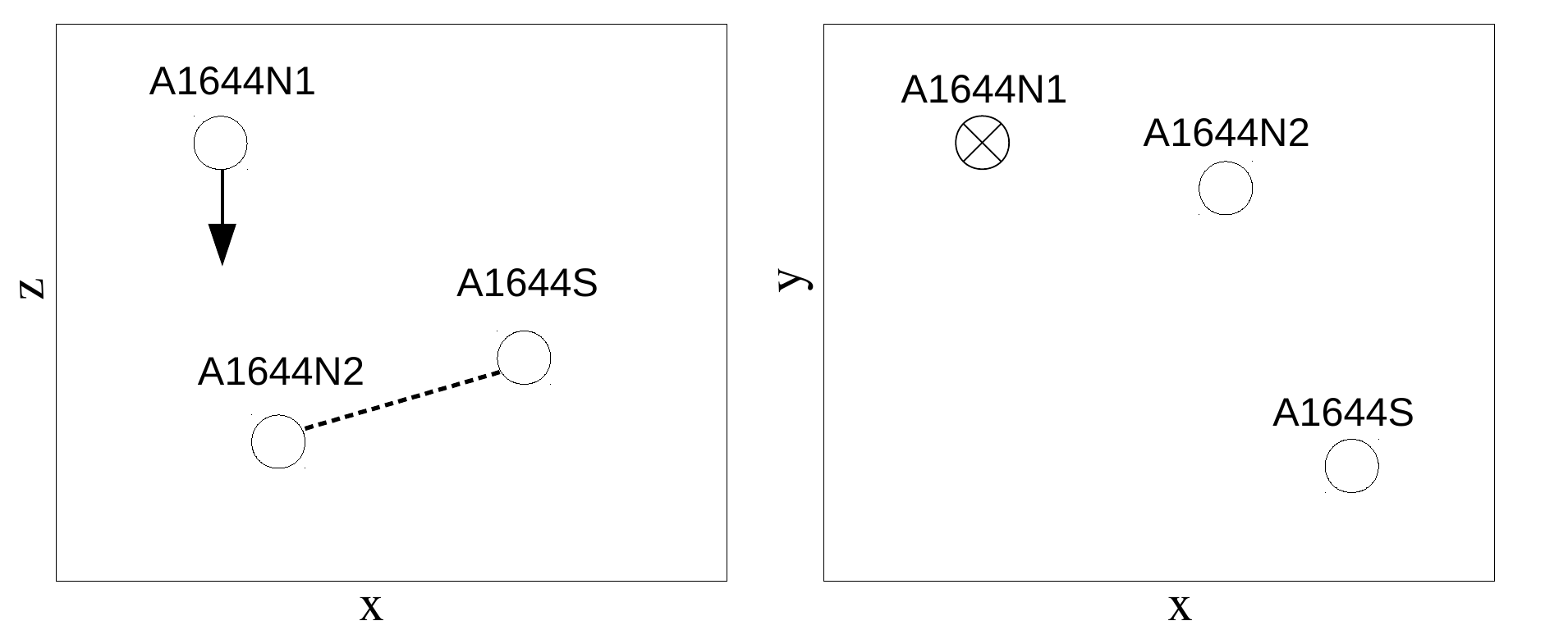}
\caption{Sketch of the three-body simulation setup. Left panel: Dotted line illustrates the inclined collision plane between A1644S and A1644N2, with the $z$-axis being the line of sight. Right panel: View representing the plane of the sky. A1644N1 arrives perpendicularly to the plane.}
\label{Fig:fig_08}
\end{figure}

\subsection {Three-body simulation} \label{3bodies}
Dedicated cluster merger simulations with more than two initial objects are relatively uncommon. For example, a triple merger has been simulated by \cite{Bruggen2012} in order to model 1RXS J0603.3+4214, and recently \cite{Ruggiero2019} simulated a quadruple merger, in an attempt to study the formation of jellyfish galaxies in the A901/2 system. In the same way, in order to reproduce the current state of A1644 with the three main bodies, we performed a simulation involving A1644S, A1644N1 and A1644N2. 

After analyzing the implications of having A1644N1 as the responsible for the sloshing phenomenon, we return to the first approach: the collision between A1644N2 and A1644S.

We choose, for simplicity, to reproduce the scenario only for model B, adding the missing galaxy cluster perpendicularly to the plane of the sky, as the simulation sketch setup presented in Fig.~\ref{Fig:fig_08}. This highly idealized approach is designed to minimize the disturbance in the plane of collision caused by the arrival of a third massive structure. In this case, we consider the collision between A1644S and A1644N2, adding A1644N1 coming through the line of sight. We are aware that this approach would lead to high LOS velocity for A1644N1, but we emphasize that in this first trial the objective is to successfully reproduce the presence of the entire system.

\begin{figure}
\includegraphics[width=\columnwidth]{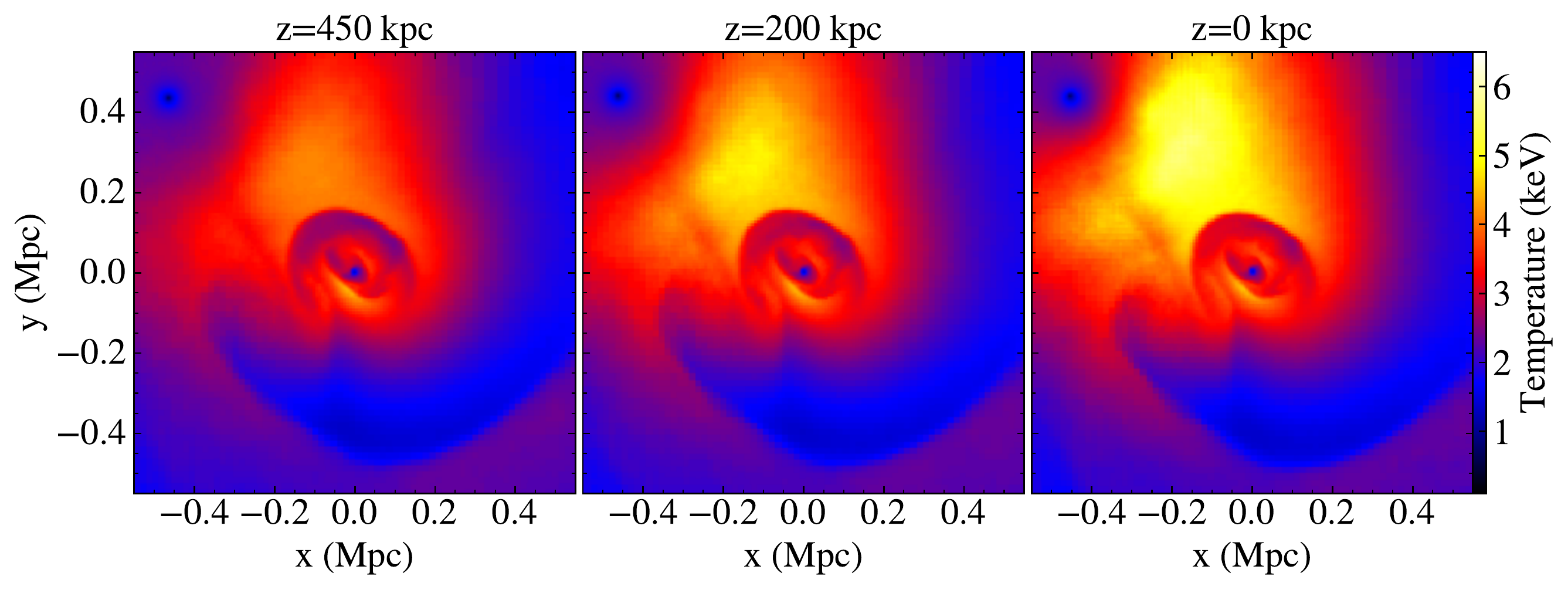}
\vspace{-10pt}
\caption{Emission-weighted temperature maps of the three-body simulation for three heights of N1 above the $xy$ plane: $450$, $200$ and $0$ kpc.}
\label{Fig:fig_09}
\end{figure}

\begin{figure*}
\centering\includegraphics[width=1.9\columnwidth]{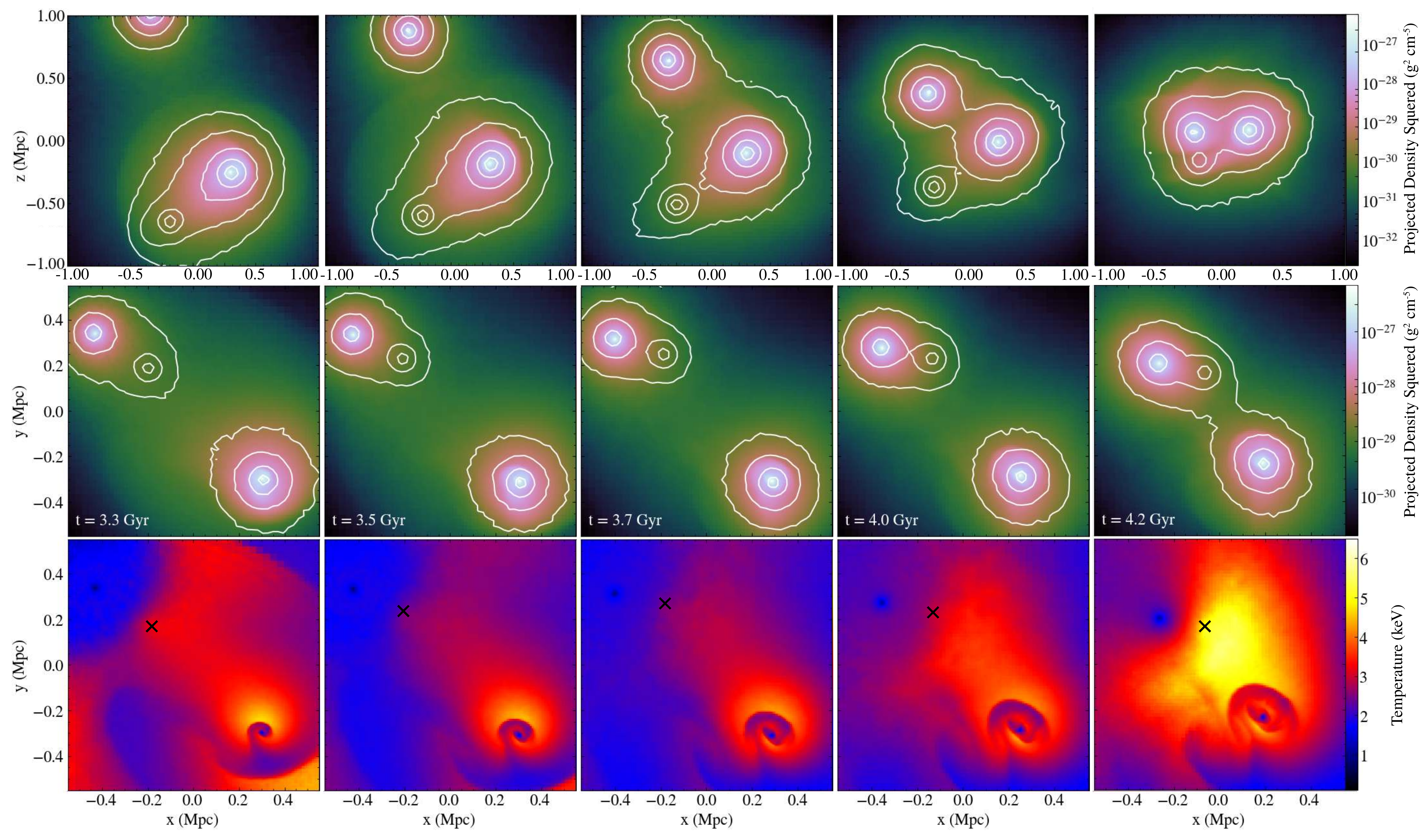}
\caption{Temporal evolution of the three-body merger simulation. Model B is the basis of the simulation, with A1644N1 arriving perpendicularly to the plane of the sky. Top panels: Projected density squared for the $x$,$z$ plane, highlighting the motion of A1644N1 along the line of sight. Middle panels: Projected density squared for the $x$,$y$ plane. Bottom panels: Emission-weighted temperature maps, with  the $\times$ symbols representing the position of the dark matter centroid of A1644N2. The arrival of A1644N1 causes an increase in the ICM temperature. Each column shows one instant.}
\label{Fig:fig_10}
\end{figure*}

\begin{figure*}
\includegraphics[width=1.9\columnwidth]{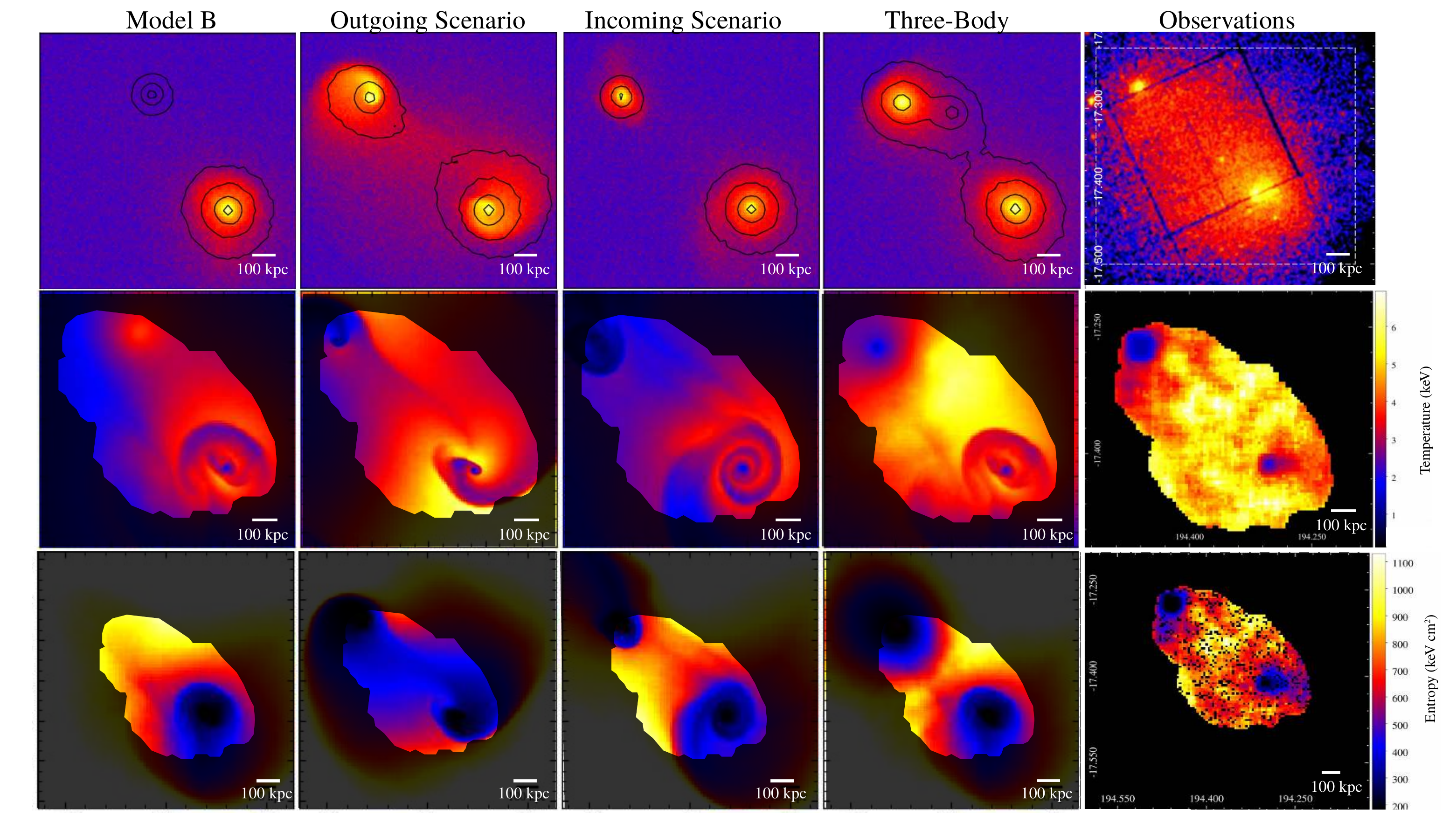}
\caption{Comparison between the best-fitting models for the galaxy cluster Abell 1644. Columns from left to right: Collision between A1644S/N2, A1644S/N1 in incoming scenario, A1644S/N1 in outgoing scenario, three-body simulation and XMM-Newton observations \citep[adapted from][]{Lagana2019}. Top panels: Mock X-ray observations from the output simulations. Black contour lines show the dark matter projected mass indicating the presence of the gas poor subcluster A1644N2. Middle panels: Emission-weighted temperature maps. Bottom panels: Entropy maps. Temperature and entropy maps were masked to highlight the region where observational data are available.}
\label{Fig:fig_11}
\end{figure*}

\begin{table}
    \centering
    \vspace{-10pt}
    \caption{Initial position coordinates and velocity vectors of A1644S, A1644N1 and A1644N2 for the triple merger. Values in relation to the center of mass of the system. Negative velocities indicate infalling motion.}
    \begin{tabular}{l c c c c c c }
        \hline
        \,     & $x$  & $y$   & $z$ & $v_x$ & $v_y$ & $v_z$  \\
        \,     &  (kpc)    & (kpc) &(kpc)  & (km/s) & (km/s) & (km/s)  \\
        \hline
        A1644N1& $-$430 &  337 & 1055 &    0 & $-$2  & $-$781     \\ 
        A1644S &  300 & $-$293 & $-$268 &   27 & $-$80 & 323     \\
        A1644N2& $-$190 & 180  & $-$640 & $-$445 & 625 & $-$140   \\
        \hline 
    \end{tabular}
    \label{Tab:Tab2}
\end{table}

In our simulations, we tried different separations between A1644S--N1 above the $xy$ plane to reproduce some inclinations of the collision plane in relation to the plane of the sky: about $30^\circ$ or $z = 450$ kpc; and the maximum approximation as both structures in the plane of the sky, $z = 0$. Structures as massive as those utilized in the collision scenarios between A1644N1 and A1644S would cause great disturbances in such small distances, yet Fig.~\ref{Fig:fig_09} shows a good representation of the overall temperature map in small displacements, when compared to the observations \citep{Reiprich2004,Lagana2019}. 

Thus, we took the case with $z = 0$ kpc as best model for further comparisons with X-ray surface brightness and thermodynamic properties. Table \ref{Tab:Tab2} summarizes the position coordinates and velocity vectors of all 3 clusters in the beginning of this simulation, that is, at $t=3.3$ Gyr.

A time evolution of the three-body simulation dynamic is presented in Fig.~\ref{Fig:fig_10}. The first two rows displays projected density squared for the ($x$,$z$) and ($x$,$y$) planes. The upper row shows the line-of-sight motion of A1644N1, arriving from $1.5$ Mpc at $t=3.2$ Gyr, and heading towards the orbital plane of A1644S/N2. The middle row exhibits, on the same time-scale, the evolution of the system in the plane of the sky. In the bottom row, the emission-weighted temperature maps show how the ICM is disturbed by the arrival of 1644N1.

Fig.~\ref{Fig:fig_11}, in the ``Three-Body'' column, shows the mock X-ray surface brightness image with contours of dark matter at the best instant, $t=4.3$ Gyr, generated as presented in Sec.~\ref{residuals} for the three-body output (a Poisson noise was added, with standard deviation proportional to the background count/pixel), alongside the temperature and entropy maps; and on the right, the observed X-ray image, temperature and entropy taken from \cite{Lagana2019}. At this instant of time the three galaxy clusters are in their desired projected positions, and the emission from A1644N2 becomes almost undetectable compared to the other two clusters. 

Even with the good agreement between $M_{200}$, $r_{200}$ and gas morphology, small scale details of the observed temperature map cannot be reproduced. This occurs, probably, due to the simplified models, that do not take substructures into account, nor other physical processes related to galaxies. However, the range of the temperature maps still shows quite good agreement.

\section{Summary and conclusions} \label{Summary}
The peculiar morphology of A1644 suggests that the galaxy cluster has undergone a recent off-axis merger. X-ray observations reveal a spiral structure with extent of $\sim200$ kpc, consistent with the sloshing phenomenon, where the gravitational pull caused by the passage of a subcluster is understood as the responsible for the spiral morphology.
%
To reconstruct the dynamical history, we use idealized $N$-body hydrodynamical simulations, focusing on the structures in a large scale, specially the spiral-like morphology of the southern cluster, so internal features ($<10$kpc) are beyond the scope of this work.

From previous studies it would be assumed, in principle, that A1644N1 was the natural disturber candidate, in a recent collision \citep{Johnson2010}. However, the gravitational lensing results \citep{Rogerio2019} revealed two massive objects that could trigger the sloshing phenomenon in A1644S: the expected northern cluster (A1644N1), a nearly undisturbed cluster that is quite visible on X-ray; and A1644N2, an gas poor subcluster nearly undetectable on X-ray.

In such circumstances of a galaxy cluster merger, it is conceivable that the disturbing subcluster has been dispersed by tidal forces, becoming impossible to identify. Following this approach, we perform a search in the parameter space, in order to restrict some of the collision parameters, taking A1644N2 as responsible for triggering the gas sloshing.

The best-fitting model must meet some requirements simultaneously, such as: the virial masses and radii for both structures, as measured in \cite{Rogerio2019}; the general morphology of the gas sloshed with an extent of approximately $200$ kpc; projected separation between peaks of dark matter of $\sim550$ kpc; and A1644N2 should present a low fraction of gas, to be nearly undetectable in X-ray observations.
 
We can not state that the best-fitting model is a unique solution, since other combinations of parameters can also present similar results. However, the ranges of values explored at least allow us to safely discard certain combinations of parameters.

Here, we summarize the four main parameters that have been constrained and provide approximate estimates of the best ranges: a) Small impact parameters are ruled out as they lead to inappropriate morphologies. A good lower limit to $b_0$ is set to $800$ kpc, which implies at pericentric passage a separation of $>180$ kpc. b) The relative velocity is constrained not only by the morphology, but also by the temperature and A1644N2 gas loss; $v=700$ km/s gives a good overall agreement. c) A1644N2 gas concentration determines essentially how bright the X-ray peak is. With scale length close to $r_{\rm b,g}=350$ kpc A1644N2 still lose sufficient gas to become undetectable in observations. d) Small inclination angles between the collision axis and the plane of the sky does not greatly affect the spiral morphology, so $30^\circ$ is adopted for consistency with the low LOS velocity between the structures discussed in \cite{Rogerio2019}.

This best-fitting model is consistent with older scenario case, where the subcluster is falling back towards the main cluster, after reaching the apocentre. In this case, the system is seen $1.6$ Gyr after the pericentric passage. 
A rough calculation from the simulation output indicates LOS velocities around $\sim$100\,km/s, in agreement with the low inclination of the plane of collision in relation to the plane of the sky.

Quantitative and qualitative comparisons with \cite{Xcop} and observational data as presented by \cite{Reiprich2004, Johnson2010, Lagana2019} were also taken into account, in the form of emission-weighted temperature maps, entropy and dark matter mass contours maps (Fig.~\ref{Fig:fig_11}), and also in profiles of thermodynamic properties (Fig.~\ref{Fig:fig_06} and Fig.~\ref{Fig:fig_07}). All the analysis presented an acceptable overall agreement.

The observed X-ray maps show two clear peaks corresponding to A1644S and A1644N1, but the X-ray counts at the position of A1644N2 are not nearly as important. In this sense, it may resemble some of the X-ray-underluminous Abell clusters studied by \cite{Popesso2007} and \cite{Trejo-Alonso2014}. With the \textit{Chandra} X-ray observations and using the {\sc Pimms} simulation tool we were able to estimate the gas mass inside $55$ kpc in A1644N2 position, $M_{\rm gas}=4.6^{+12.8}_{-4.0} \times 10^{10}~{\rm M}_\odot$. 
While the measured mass from the simulations is $M_{\rm gas}=1.4 \times 10^{10}~{\rm M}_\odot$. These results suggest the the gas fraction of A1644N2 should be somewhere in the range of only 0.1--1 per cent.

So far, we have presented our best collision scenario for A1644, the collision between A1644S and A1644N2. However, until recently, the most obvious candidate would have been A1644N1. So we also explored the possibility of a collision between A1644S and A1644N1.
These scenarios did not give rise to results as satisfactory as the model B. In a first passage, in the outgoing scenario, the shock waves are still too evident, rising the ICM temperatures up to $\sim6$ keV in the vicinity of the spiral feature, and also great disturbances in the density map are seen in both clusters cores.  After reaching the maximum separation, A1644N1 starts to fall back in the direction of A1644S, in this incoming scenario, the ICM becomes too cold ($<3.5$ keV) at the best spiral-like morphology instant.

As the number of constraints increases, it becomes progressively more challenging to obtain a model that accommodates them all simultaneously and quantitatively. This is one of the reasons why tailored simulations of collisions involving more than two clusters are seldom attempted.
Still, as a tentative complementary simulation, we added A1644N1 arriving perpendicularly to model B. Prominent disturbances were expected in the morphology of the system, however this attempt to reproduce the presence of the three major structures showed a good agreement to the global morphology of the observations. Similarly, when S and N1 are in the plane of the sky the  simulated emission, temperature and entropy maps are within the expected range presented by \cite{Lagana2019}.

We conclude that for the two possible scenarios that may explain the current morphology of gas sloshing in the southern cluster of A1644, the one that appears to give the best fitting model is the collision between A1644S and A1644N2. A1644N1 may be present as long as it does not interfere in the formation of the spiral feature. The solution where A1644N1 is the disturber is difficult to find, since the collision between two clusters with similar masses leads to great disturbances. Thus, we find that the more likely scenario is that of a collision with a lower-mass subcluster that loses its gas after the pericentric passage.

\section*{Acknowledgements}

This work made use of the computing facilities of the Laboratory of Astroinformatics (IAG/USP, NAT/Unicsul), whose purchase was made possible by the Brazilian agency \textit{Funda\c c\~ao de Amparo \`a Pesquisa do Estado de S\~ao Paulo} (FAPESP - grant 2009/54006-4) and \textit{Instituto Nacional de Ci\^encia e Tecnologia em Astrof\'isica} (INCT-A). Simulations were carried out in part at the \textit{Centro de Computa\c c\~ao Cient\'ifica e Tecnol\'ogica} (UTFPR). REGM acknowledges support from the Brazilian agency \textit{Conselho Nacional de Desenvolvimento Cient\'ifico e Tecnol\'ogico} (CNPq) through grants 303426/2018-7 and 406908/2018-4. TFL acknowledges financial support from the Brazilian agencies FAPESP and CNPq through grants 2018/02626-8 and 303278/2015-3, respectively. GBLN acknowledges support from CNPq and FAPESP (grants 306498/2010-3 and 2018/17543-0). RMO acknowledges the financial support provided by \textit{Coordena\c{c}\~{a}o de Aperfei\c{c}oamento de Pessoal de N\'ivel Superior} (CAPES). ESC acknowledges support from the CNPq through grant 308539/2018-4. This study was financed in part by CAPES - Finance Code 001. We thank the referee for the helpful comments that improved the paper.

\bibliographystyle{mnras}
\bibliography{simulacao.bib} 

\bsp	
\label{lastpage}
\end{document}